\documentclass[aps,prx,twocolumn,superscriptaddress]{revtex4}
\usepackage{graphicx}
\usepackage{amssymb}
\usepackage{amsmath}
\usepackage{epsfig}
\usepackage{color}
\usepackage{float}
\usepackage{threeparttable}
\usepackage{booktabs}
\usepackage[colorlinks=true,
			linkcolor=blue,
			urlcolor=cyan,
			citecolor=blue]{hyperref}
\usepackage{hyperref}
\usepackage{lipsum}		
\usepackage{ulem}
\begin{document}
\title{Spectroscopic data de-noising via training-set-free deep learning method}
\author{Dongchen Huang }
\affiliation{Beijing National Laboratory for Condensed Matter Physics and Institute of
Physics, Chinese Academy of Sciences, Beijing 100190, China}
\affiliation{University of Chinese Academy of Sciences, Beijing 100049, China}

\author{Junde Liu }
\affiliation{Beijing National Laboratory for Condensed Matter Physics and Institute of
Physics, Chinese Academy of Sciences, Beijing 100190, China}
\affiliation{University of Chinese Academy of Sciences, Beijing 100049, China}

\author{Tian Qian}
\email[Corresponding author:]{tqian@iphy.ac.cn}
\affiliation{Beijing National Laboratory for Condensed Matter Physics and Institute of
Physics, Chinese Academy of Sciences, Beijing 100190, China}
\affiliation{University of Chinese Academy of Sciences, Beijing 100049, China}
\affiliation{Songshan Lake Materials Laboratory, Dongguan, Guangdong 523808, China}
\author{Yi-feng Yang}
\email[Corresponding author:]{yifeng@iphy.ac.cn}
\affiliation{Beijing National Laboratory for Condensed Matter Physics and Institute of
Physics, Chinese Academy of Sciences, Beijing 100190, China}
\affiliation{University of Chinese Academy of Sciences, Beijing 100049, China}
\affiliation{Songshan Lake Materials Laboratory, Dongguan, Guangdong 523808, China}
\date{\today}
\begin{abstract}
    De-noising plays a crucial role in the post-processing of spectra. Machine learning-based methods show good performance in extracting intrinsic information from noisy data, but often require a high-quality training set that is typically inaccessible in real experimental measurements. Here, using spectra in angle-resolved photoemission spectroscopy (ARPES) as an example, we develop a de-noising method for extracting intrinsic spectral information without the need for a training set. This is possible as our method leverages the self-correlation  information of the spectra themselves. It preserves the intrinsic energy band features and thus facilitates further analysis and processing. Moreover, since our method is not limited by specific properties of the training set compared to previous ones, it may well be extended to other fields and application scenarios where obtaining high-quality multidimensional training data is challenging.
\end{abstract}

\maketitle

\section{Introduction}\label{section1}
		
	Recent developments of experimental techniques in physics research have facilitated the generation of large amounts of high-resolution, high-dimension, and high-complexity data. As a result, one may often need to analyze and process two or even three-dimensional spectra, such as in angle-resolved photoemission spectroscopy (ARPES) \cite{Damascelli2003,Lv2021,Sobota2021}, scanning tunneling microscopy (STM) \cite{Binnig1983,Pan2001a}, inelastic neutron scattering (INS) \cite{Chaix2013a}, resonant inelastic x-ray scattering (RIXS) \cite{Kotani2001a}, and momentum-resolved photoemission electron microscopy (k-PEEM) \cite{Sobota2021,Medjanik2017a}. The enhanced resolution of these spectra enables the identification of fine-grained structures that may lead to the discovery of new phenomena.

    In spectroscopic experiments, high-quality data are of critical importance for extracting important detailed information. But a good signal-to-noise ratio (SNR) level often requires a long  spectrum acquisition time, which is difficult to achieve in many situations. For instance, the time of ARPES measurement is typically limited because of the shortage of synchrotron light resources or the ageing of sample surfaces due to the adsorption of remaining gas molecules. Although a shorter acquisition time might be achieved by increasing the light intensity or changing the analyzer settings, it also reduces the resolution and brings up a series of other problems, such as space charge effects \cite{Zhou2005a,Graf2010a}, detector non-linearity effects \cite{Smallwood2012b,He2016a}, or photo-induced sample damage \cite{Mills2019a,Grass2010a,Nitta2019a}. Therefore, post-processing such as de-noising is usually necessary for analyzing high-dimensional spectra with low SNR levels.
    
    At present, noise reduction processing for spectroscopic data is mostly based on mathematical methods such as Gaussian smoothing and Fourier transform filtering. These methods are often not very effective and may lose certain intrinsic information due to over-processing. Recently, machine learning and deep learning methods based on convolutional neural networks (CNN) have been rapidly developed for spectra processing to achieve super-resolution \cite{Peng2020a} and de-noising \cite{Kim2021a}, or solving more general condensed-matter-physics problems \cite{Huang2022,Dong2021}. However, all the existing de-noising methods require a sufficiently large training set, or it may easily work outside the training domain and lose robustness in practice \cite{Antun2020}. We may refer to this phenomenon as a hallucination, which has been reported in the medical image processing  \cite{Bhadra2021}. For image, it has been shown that using self-correlation may avoid the training set with the noisy image as the sole input \cite{Li2013a,Ulyanov2020}.

    In this work, we point out that spectroscopic data in condensed matter physics are also self-correlated and may well be treated by applying similar techniques in image processing. We demonstrate that the training-set-free de-noising method also performs well for ARPES spectra to extract intrinsic information with the help of deep learning. Compared with previous deep-learning-based de-noising methods, our proposed method avoids the collection of training sets and has higher adaptability and flexibility. It may thus be applied to more general scenarios where a good training set is hard to obtain.

\section{Methods}\label{sec:2}
    \begin{figure*}[t]
    \centering
    \includegraphics[width=1.0\textwidth]{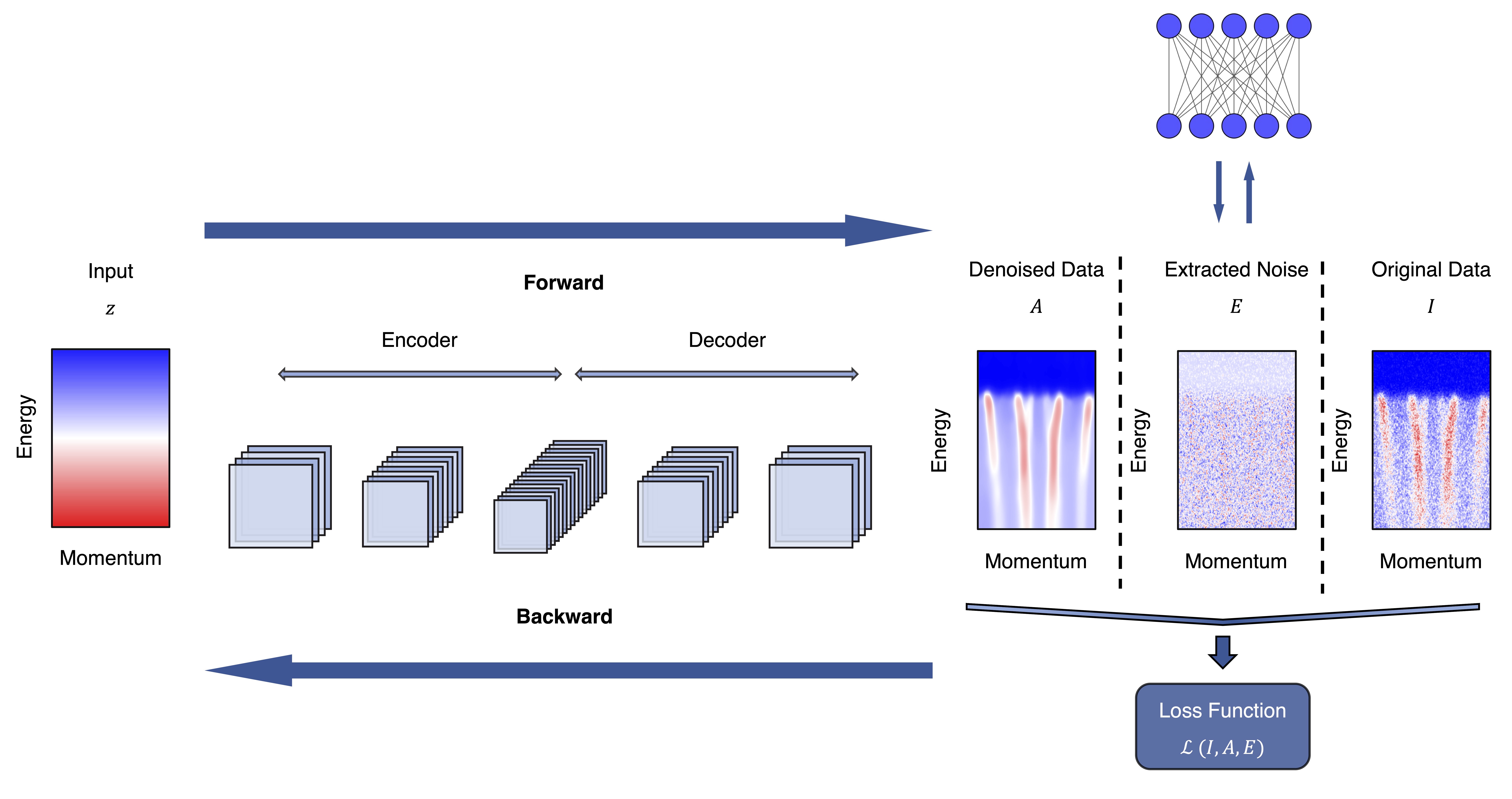}
    \caption{A schematic plot of the parameterization and encoder-decoder framework. The clean spectra and the noise are parameterized by two independent neural networks. The clean spectra is parameterized by the encoder-decoder network, while the noise is parameterized by a small network. The encoder receives input and maps it to the latent space, while the decoder maps the latent space back to the input space. These two neural networks first guess a superposition of clean spectrum and noise via forward propagation, and then the parameters of the neural network are updated with respect to the loss function through backpropagation (BP). Both the encoder and decoder are convolutional neural networks (CNNs) containing five convolutional layers, and the noise network is introduced in the main text.}
    \label{Fig1}
    \end{figure*}
	We consider a noisy ARPES image whose $x$- and $y$-axes denote momentum and energy, respectively. Our method is based on the decomposition of the image into two parts: a correlated part containing the desired spectral information and a noise part which is presumably less correlated with ground-truth spectra. Such decomposition is reasonable for representing the noisy ARPES image in which nearby columns and rows are highly correlated while the distribution of noise is somewhat diverse.

    Mathematically, let $I \in \mathbb{R}^{m\times n}$ be the noisy ARPES image, where $m$ and $n$ are the number of discretized energy and momentum, respectively. We aim to find a decomposition that takes the form: 
    \begin{equation}
	    I = A + E,
    \end{equation}
    where $A$ is the desired clean ARPES image and $E$ denotes the noise.

    \paragraph{Neural Network and Convolutional Neural Network} We use neural networks to achieve the clean image $A$ and the noise $E$. An general $L$-layer neural network is a composition of $L$ non-linear mapping and can be seen as a function of the form:
    \begin{equation}
        F(x) = f^L \circ f^{L-1} \circ \cdots \circ f^1(x),
    \end{equation}
where $x$ is any given input and $f$ represents an individual layer composed of two maps: a linear mapping and a usually element-wise non-linear function to enhance the nonlinearity.

    The convolutional neural network (CNN) is a special architecture enabling the explicit modelling of local correlation by utilizing convolutional layers, where the linear feature mapping is no longer a general matrix multiplication but a convolution operation between the input single and a convolution kernel.  For example, for the $l$-th convolution layer, it has the form:
    \begin{equation}
        f^l(x) = \phi(w * x + b),
    \end{equation}
    where the input signal $x$ is assumed to be 1D for simplicity, the weight $w$ is called the convolution kernel and $b$ denotes the bias. To deal with 2D or 3D spectra, it is necessary to replace the 1D convolution by 2D or 3D ones. $\phi(.)$ is an element-wise function and stands for the activation function. It is usually chosen to be the Leaky Rectified Linear Unit (Leaky ReLU), which takes the form:
        \begin{equation}
            \phi(x)= \left\{\begin{array}{ll}
        x \quad &x>0 \\
        \alpha x & x\leq 0
    \end{array}\right. 
    \end{equation}
    where $\alpha\geq 0$ is a small constant. If $\alpha = 0$, the Leaky ReLU function reduces to another commonly used function called Rectified Linear Unit (ReLU).

    The feature size can be reduced by introducing downsampling layer, and two common techniques are widely applied. One is max-pooling, which is done by applying a max filter to non-overlapping sub-regions of the feature, and the other is strided convolution, which can be done by changing the stride in convolution operation.
    
    \paragraph{Parameterization by deep neural network} 
    A natural question is how to find a correct parameterization scheme for the clean image $A$ and the noise $E$. Fortunately, recent advances in deep learning suggest that deep neural networks are capable of achieving great success \cite{Krizhevsky2017,Brown2020} in modelling correlated data like images and languages. Especially, convolutional neural networks can enforce structural and correlated priors. Motivated by these successful practices, we parameterize the correlated part $A$ as the output of a deep neural network using the encoder-decoder framework. This architecture is practical for modelling such data \cite{Ronneberger2015UNet}. A schematic diagram of the encoder-decoder framework is shown in \ref{Fig1}, which contains two symmetric networks: the encoder and the decoder. The encoder network has five convolutional layers to model the local connection and map the input image to the latent space. The decoder network contains five deconvolutional layers, but plays the opposite role and maps the data from the latent space to the input space.
    
    The parameterization of the noise $E$ is less obvious. We assume that the noise is sparse providing that not all pixels of the noisy ARPES image are heavily corrupted. Following the recent advances in non-convex optimization \cite{You2020}, any sparse vector $e$ may be parameterized as $g\circ g - h\circ h$ where $g,h$ are two vectors and $\circ$ denotes their element-wise product.

    Taken together, for a given noisy ARPES image $I$, we need to minimize the following loss function:
    \begin{equation}
	    \mathcal{L} = \|{A_\theta + g\circ g - h\circ h}-I\|_2^2,
	    \label{Eq:loss}
    \end{equation}
    where $\|.\|_2^2$ denotes the $\ell^2$ norm \footnote{The $\ell^2$ norm of a vector is the sum of the square of every entry.}, $A_\theta$ is the desired de-noised spectra flattened as a vector, $\theta$ is the collection of parameters of the neural network, and $g,h$ represent the sparse noise $E$. 
    
    \paragraph{Optimization} 
    As illustrated in ~Fig. \ref{Fig1}, it is necessary to iterate many times for the neural network to find the proper parameters. In each iteration, the neural network first gets the input and guesses both the clean spectral image and the noise, whose quality is then evaluated by the loss function given in eq. \eqref{Eq:loss}. The parameters are then updated for the next iteration according to the loss function. This may be called the training process. The input can be somewhat arbitrary, and the only requirement might be the reflection of self-correlation. We choose the input following the practice reported in \cite{Ulyanov2020}. However, the input cannot be the original noisy data. Otherwise, the global minimum of the loss function \eqref{Eq:loss}, which is zero, would be achieved at $g = h = 0$, so the neural network learns nothing but the identity map. This is different from other algorithms which can have access to the ground-truth high-quality data, where the neural network is trained to fit the noiseless high-quality spectra. By contrast, we only have the noisy spectra and train the neural network to fit the noisy spectra of low quality. In practice, we can also choose other kind of inputs. After training, the neural network can output both the desired clean spectra and the noise. 
    
    All parameters are trained via stochastic gradient descent (SGD) with discrepant learning rates: $\eta_a=1$ for the encoder-decoder network and $\eta_e=2500$ for the noise. The discrepant learning rate $\eta_e$ plays a crucial role in the performance of our algorithm, which will be discussed later. Our method is implemented by Pytorch \cite{Paszke2019} and more details are given in Appendix A2.

\section{Applications}\label{sec:3}
	
\subsection{Two-dimensional data}

	\begin{figure*}[ht]
	\centering
	\includegraphics[width=0.8\textwidth]{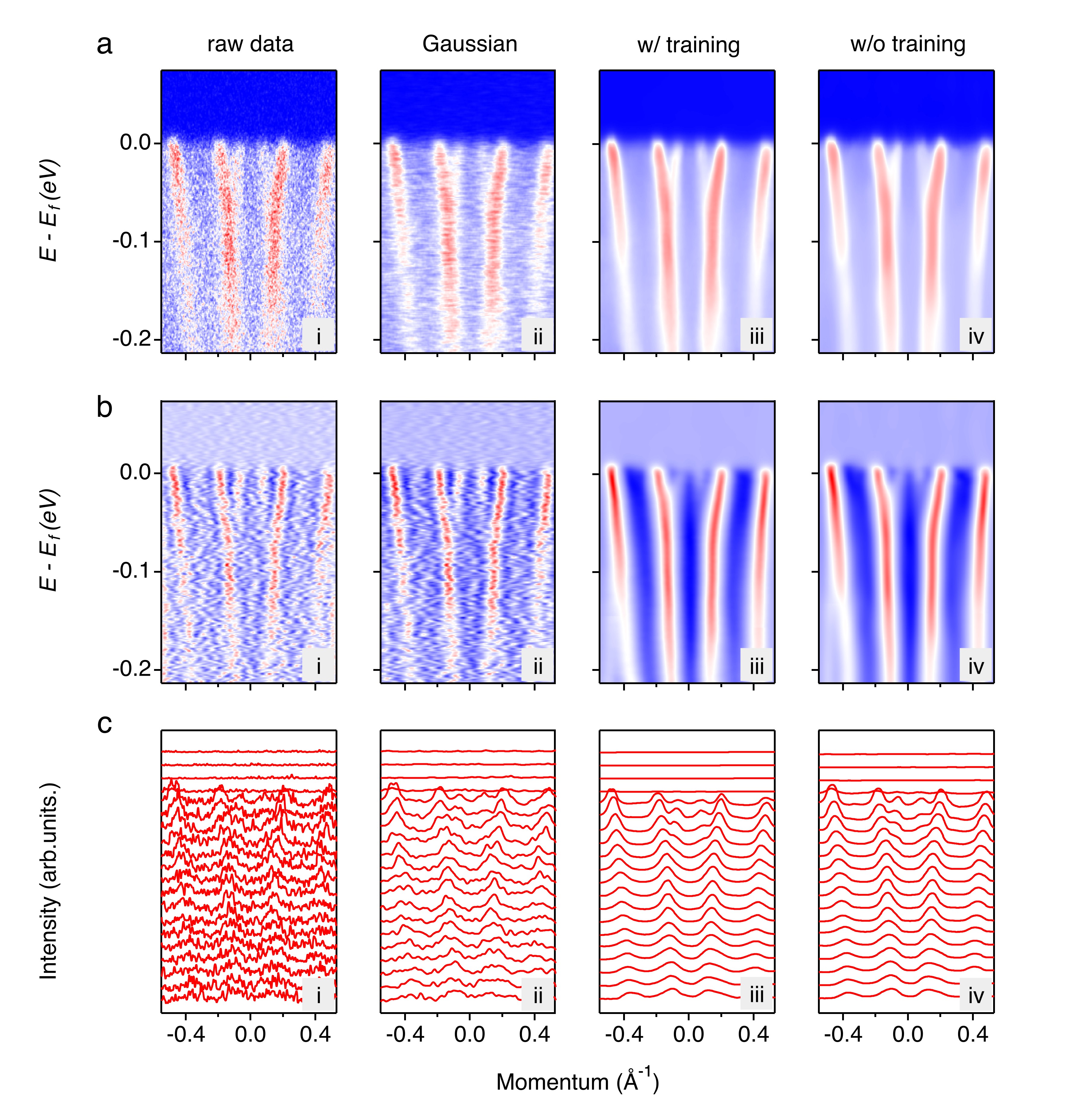}
	\caption{Comparison of the de-noised results by different approaches. (a) ARPES intensity spectra in Bi2212 along the nodal cut. (i) The noise-corrupted raw data used for the following de-noising process. (ii-iv) The de-noised results using (ii) Gaussian smoothing method, (iii) convolutional neural networks method with training set, and (iv) convolutional neural networks method without training set. (b) Second-derivative plots from the corresponding data in (a). (c) Momentum distribution curves (MDCs) from the corresponding data in (a).}
	\label{Fig2}
    \end{figure*}
    As an example, we apply our method to the ARPES spectra of Bi2212 along the nodal cut. The intensity plots of the original data and those after de-noising using different methods are compared in Fig. \ref{Fig2}(a), where we compare our method with both Gaussian smoothing and the training-set based deep learning method with the same training set as proposed in \cite{Kim2021a}. The original data contain a high level of noise due to a short acquisition time. Moderate Gaussian smoothing can only attenuate part of the high-frequency noise. As a result, the overall spectra remain not so smooth, but further increasing the number of smoothing operations will blur the spectra, smear out the band features, and lose some fine details. Therefore, the Gaussian smoothing method is not effective for removing sparse noise. By contrast, both deep learning methods, whether a training set is required \cite{Kim2021a} or not, can both perform well. The noise can be successfully removed from the spectra while preserving the dispersive features of the energy bands. Our training-set-free method achieves comparable results as the training-set-based deep learning algorithms. Moreover, because of the good performance, the diverse and weak correlation assumption is found to be no longer a bottleneck of our method in this application.

    The second derivative of the data is plotted in Fig. \ref{Fig2}(b) for better visualization of the band structures, which is often used in ARPES experiments. Again, the quality of the energy band dispersion is neither very clean nor smooth in the original spectra Fig. \ref{Fig2}(b-i), and the Gaussian smoothing method shows no significant improvement either. By contrast, the second derivative of our de-noised data exhibits very clear and smooth band dispersions, in particular the well-known 70 meV kink and superlattice feature, which are not such clearly resolved in the noisy raw data. Since the second derivative is very sensitive to noise, removing the noise has a great impact on the performance. Therefore, better de-noising is very  helpful for a clear visualization of the energy bands, which demonstrates the usefulness of deep-learning-based de-noising methods for post-processing.

    Fig. \ref{Fig2}(c) further shows the momentum distribution curves (MDCs) of the spectra for a more intuitive presentation of the performance of our de-noising method. The raw data  presented in Fig. \ref{Fig2}(c-i) are so noisy that the peak shape can barely be identified. The Gaussian smoothing can only remove the high-frequency noise but retains the low-frequency noise in the MDCs. As shown in Fig. \ref{Fig2}(c-ii), the curves are still noisy. More seriously, as the number of smoothing increases, the peak width also increases, thus losing the intrinsic information of the energy bands and causing difficulties for further quantitative analyses such as self-energy extraction. By contrast, the de-noised results by the deep-learning method in Fig. \ref{Fig2}(c-iii) and Fig. \ref{Fig2}(c-iv) are very smooth, and the intrinsic features of the band structures are well preserved, including the peak position and width. More examples are given in Appendix A1 that confirm the robustness of our method.

\subsection{Three-dimensional data}
	\begin{figure*}[ht]
	\centering
	\includegraphics[width=0.8\textwidth]{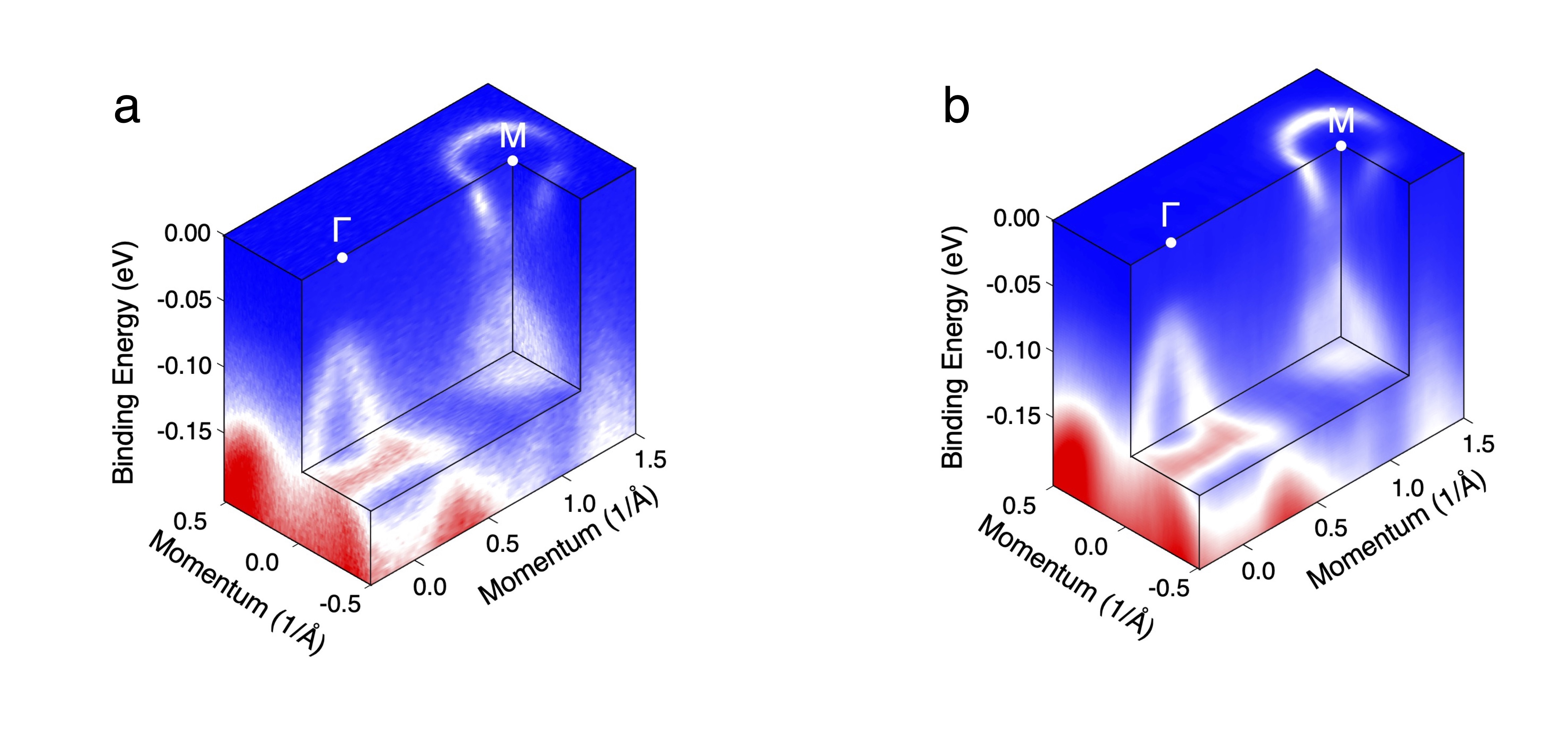}
	\caption{Three-dimensional band mapping of the $\rm{FeSe/SrTiO_{3}}$ thin film along $\Gamma-M$ direction. (a) The noise-corrupted original data. (b) The de-noised results  based on the original data in (a) using our method with no training set.}
	\label{Fig3}
	\end{figure*} 
    For ARPES in condensed matter physics, three-dimensional data are crucial and more informative for studying the Fermi surface properties. However, previous deep learning-based de-noising methods cannot handle such three-dimensional data well because they require a large amount of expensive high-quality training data. By contrast, our training-set-free method only relies on the self-correlation of three-dimensional spectra themselves to remove the noise and can be easily extended to the three-dimensional case. Fig. \ref{Fig3} shows a  typical three-dimensional intensity plot of the $\rm{FeSe/SrTiO_{3}}$ thin film. Both band dispersion and the isoenergetic contour are clearer and smoother after the de-noising process. Thus, applying our method can indeed de-noise the three-dimensional data and enhance their SNR levels. We therefore expect that it may greatly reduce the measurement time, which is of particular importance in higher-dimensional experiments.

\section{Discussion}\label{sec:4}

	There are two key parameters in our algorithm: the noise learning rate $\eta_e$ and the number of iteration steps. In this section, we discuss the influence of the noise learning rate $\eta_e$ and the training dynamics during the iterations.
    
\subsection{Effect of the learning rate}
\begin{figure}[H]
    \centering
    \includegraphics[width=0.47\textwidth]{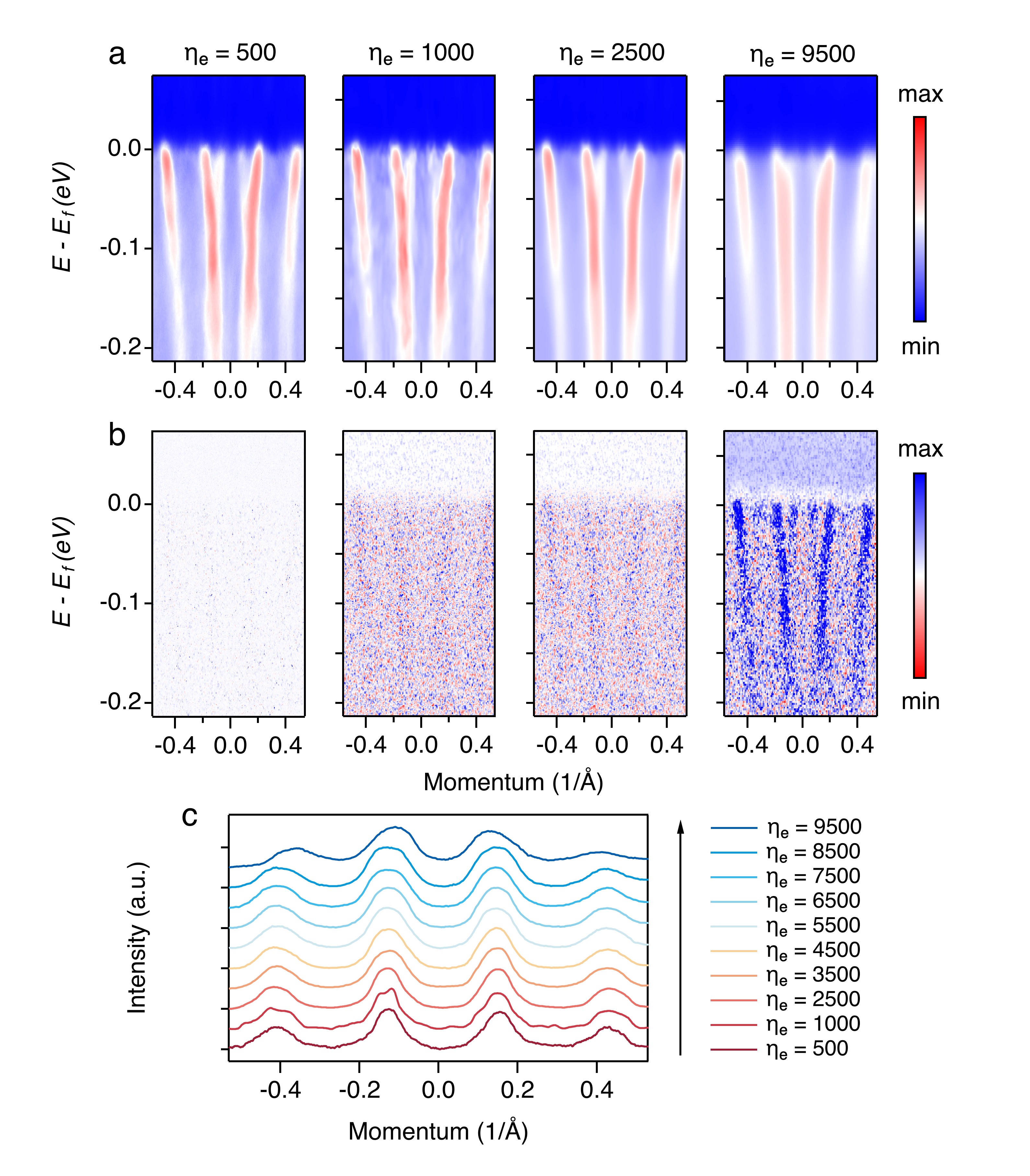}
    \caption{Comparison of the de-noised Bi2212 spectra along the nodal cut under different noise learning rates.  (a) The de-noised results for $\eta_e$=$500$, $1000$, $2500$, $9500$. (b) The corresponding noise of panel (a). (c) Variation of the MDCs at $E_B=-0.1eV$ under different noise learning rates.}
    \label{Fig4}
    \end{figure}

    The noise learning rate $\eta_e$ determines the smoothness of the de-noised ARPES image and the density of the recognized noise. It can be seen in Fig. \ref{Fig4}(a) that the de-noised images are rough for relatively lower learning rates ($\eta_e=500$, 1000) and become very smooth at a high rate of $2500$. The predicted noise in Fig. \ref{Fig4}(b) is also rather sensitive to the value of $\eta_e$. It is almost zero at every pixel but becomes denser with increasing learning rate. 
    
    Hence, a low noise learning rate creates a rough de-noised image and uniform noise around zero, while a high learning rate results in a smooth and clean spectral image but with a dense and non-uniform noise of large magnitude. These suggest a subtle trade-off mechanism. As shown in the right panel of Fig. \ref{Fig4}(a) and Fig. \ref{Fig4}(b) for $\eta_e=9500$, the resulting energy bands are over-smoothed and the predicted noise contains too much information about the intrinsic energy bands. Moreover, the MDCs illustrated in Fig. \ref{Fig4}(c) also indicate that this ultra large $\eta_e$ leads to a result with undesirable peak width and position. Clearly, a proper value of the learning rate should be chosen in order to obtain a clean spectra and at the same time avoid over-smoothing.
    
    We note that as shown in Fig. \ref{Fig4}(b), the noise distribution may not be a simple Poisson distribution but should be correlated with the distribution of the spectral intensity. Hence, the sparsity assumption of noise may not always hold. But our results suggest that this is not a bottleneck here. In fact, the real noise and the signal are always coupled together, and it is possible to extract the real noise structure without making assumptions besides sparsity.

    The effect of noise learning rate $\eta_e$ may be characterized by the following simplified model, where the non-linear autoencoder is replaced by a linear network. In this case, we can use two matrices $U$ and $V^T$ to represent the encoder and decoder networks, respectively. The optimization thus follows:
    \begin{equation}
        \min_{U,V,g,h} \mathcal{L} = \|UV^T + g \circ g - h\circ h - I\|_2^2,
        \label{Eq:Linear Model}
    \end{equation}
    where the desired ``clean spectra" now becomes $UV^T$. If discrepant learning rates are used for the "clean spectra" part $L=UV^T$ and the "noise part" $S=g \circ g - h\circ h$, we obtain 
    \begin{align}
        U_{k+1} &= U_k - \eta_a\frac{\partial \mathcal{L}}{\partial U}, \quad 
        V_{k+1} = V_k - \eta_a\frac{\partial \mathcal{L}}{\partial V}, \\ 
        g_{k+1} &= g_k - \eta_e \frac{\partial \mathcal{L}}{\partial g}, \quad  
        h_{k+1} = h_k - \eta_e \frac{\partial \mathcal{L}}{\partial h},
    \end{align}
    where $k$ is the iteration step. It has been shown that the solution of the above linear model is identical to the following convex problem \cite{You2020}, which allows for global optimization:
    \begin{equation}
        \min_{L,S} \|L\|_* + \lambda \|S\|_1 \quad s.t. \quad  L + S = I,
        \label{Eq: RPCA}
    \end{equation}
    where $\|L\|_*$ is the nuclear norm (defined as the summation of singular values) of $L$, the $\|S\|_1$ denotes the $\ell^1$ norm of the matrix $S$ (defined as the summation of the absolute value of each entry of $S$), and $\lambda=\eta_a/\eta_e$. In practice, we have set $\eta_a=1$ and $\eta_e=2500$. According to the theory of high-dimensional data analysis \cite{Candes2011}, the nuclear norm term encourages the low-rank solution. For ARPES spectra, the low-rankness models the simplest linear self-correlation among the energy bands. 

    The physical meaning of the noise learning rate $\eta_e$ is then rather clear in the above equivalent formulation \eqref{Eq: RPCA}. If $\lambda$ is small or $\eta_e$ is large, the second term in eq. \eqref{Eq: RPCA} has little effect, and the solution shows low-rankness without considering the sparse structure of the measurement noise, so the result is oversmoothing. In another word, some details of the energy bands may be ignored due to the low-rankness. On the other hand, if $\lambda$ is very large or $\eta_e$ is small, the resulting $L$ has high rank and the energy bands show roughness due to the inclusion of some undesired features related to the measurement noise. Thus,  this simple linear model captures the main behavior in Fig. \ref{Fig4} and shows that the noise learning rate $\eta_e$ should be neither too high nor too low.

     It should be pointed out that we cannot directly measure the performance of our algorithm or even rigorously define what overfitting is since we cannot access the non-existent ground truth image. On the other hand, we could view overfitting as fitting the undesirable noise in the de-noised image, which are heavily influenced by the noise learning rate $\eta_e$ and the noise parameterization. We may refer the roughness of the output due to the low noise learning rate as an indicator of overfitting. The oversmoothing output may be seen as a sign of overkill or underfitting because some physical details are ignored by the algorithm. The noise learning rate $\eta_e$ plays the role of regularization because a higher learning rate can alleviate overfitting to the noise and encourage the neural network to output a smooth image. Obviously, it should not be too high to prevent oversmoothing or underfitting. The noise parameterization is an essential part in stabilizing the neural training and avoiding overfitting. Without this part, our algorithm reduces to the deep image prior (DIP) \cite{Ulyanov2020}. However, this method will eventually result in overfitting because a neural network is directly trained to fit the noisy image. It is then necessary to apply tricky regularization techniques like early stopping \cite{Ulyanov2020}, but early stopping may be infeasible for scientific image processing because of the inaccessibility of the ground truth.
    
\subsection{Training history}

\begin{figure}[H]
	\centering
	\includegraphics[width=0.47\textwidth]{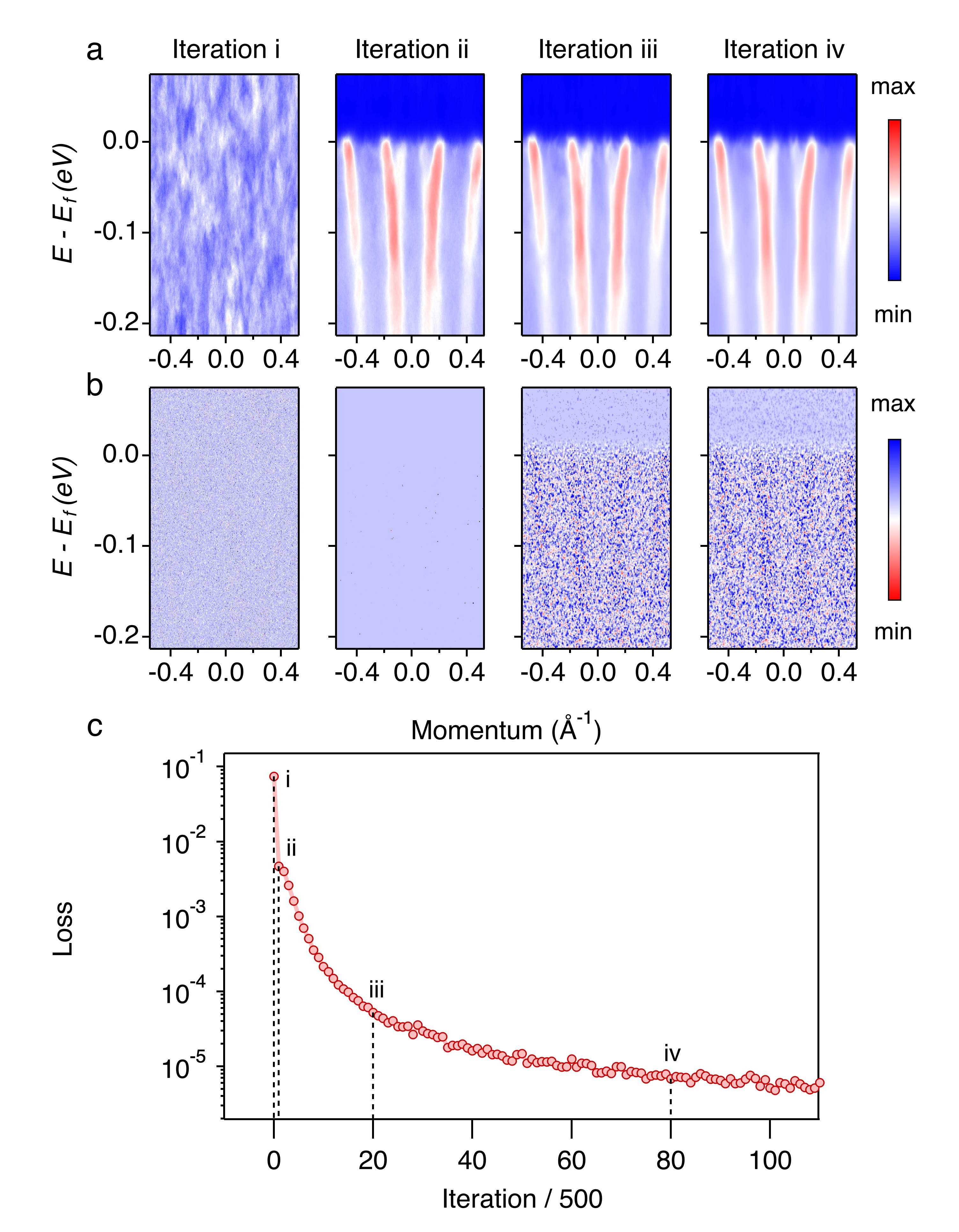}
	\caption{De-noised results of Bi2212 along the nodal cut under different iteration steps and the training curve at different stages. (a) de-noised results under four checkpoints corresponding to the iteration steps $0$, $500$, $10000$, and $40000$. (b) The corresponding noise of panel (a). (c) Validation loss as a function of the iteration number. The loss decreases quickly and the quality of de-noised images increases as well. After around $5000$ iterations, the loss already approaches zero. After $10000$ iterations, the loss remains almost constant, but the de-noised images become smoother.}
	\label{Fig5}
	\end{figure}

    The de-noised results and training curve at different stages of the training are shown in Fig. \ref{Fig5}. The neural network captures gradually the main features of the energy bands in Fig. \ref{Fig5}(a) and the noise in Fig. \ref{Fig5}(b) along the training process. The training process can thus be divided into two stages. The first stage may be viewed as feature-fitting. In the beginning, the neural network does not learn any information about energy bands or noise. After $500$ training iterations, the neural network starts to capture the main intrinsic features of the spectra, but does not extract any information about the noise. The second stage is noise-fitting and fine-tuning. The neural network begins to learn and correctly predict the noise structure. The de-noised image becomes smoother and the details of the energy bands become clearer. This phenomenon suggests the necessity of sufficient iterations. As may be seen from the loss function in Fig. \ref{Fig5}(c), our algorithm remains stable after $50000$ iterations. In fact, we have tested it up to $150000$ iterations and find it unlikely to crash.

\section{Conclusion}\label{sec:5}
	In this work, we propose a training-set-free method to handle noisy spectra effectively in both two and three dimensions. Different from other prevailing de-noising approaches, our method trains a neural network to fit a single noisy image and does not require extra training data to extract intrinsic dispersive features of the spectra. This greatly lowers the cost of collecting training sets and also avoids the hallucination problem. Our algorithm can be easily extended to other spectroscopic measurements and help with the discovery of novel spectral features in condensed matter systems.
	
\section*{ACKNOWLEDGEMENTS}
We thank Mojun Pan, Famin Chen, Jierui Huang, Bei Jiang, Mingzhe Hu for useful discussions, and Xun Shi, Yigui Zhong, Hang Li for data support. This work was supported by the National Natural Science Foundation of China (NSFC Grants Nos. 11974397, U1832202, 11888101), the Chinese Academy of Sciences (Grant Nos. QYZDB-SSW-SLH043, XDB33000000, and XDB28000000), the Informatization Plan of Chinese Academy of Sciences (Grant No. CAS-WX2021SF-0102), and the Synergetic Extreme Condition User Facility (SECUF).



\section*{\label{sec:level1} APPENDIX A: More Examples}
\begin{figure}[H]
	\centering
	\includegraphics[width=0.48\textwidth]{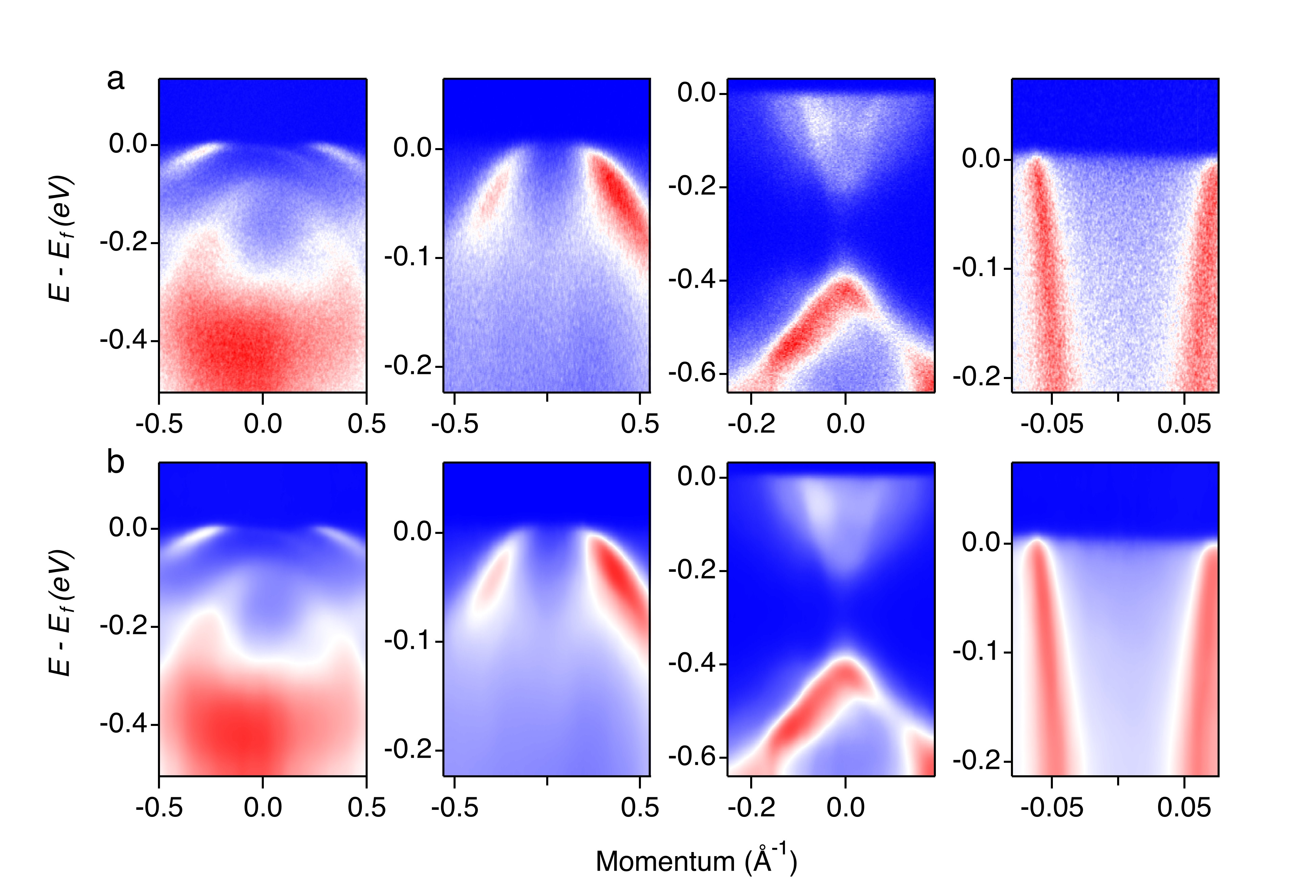}
	\caption{More examples of de-noised ARPES spectra. (a) The noise-corrupted raw data. (b) The de-noised result based on the raw data in (a) with our convolutional neural network method without training set.}
	\label{Fig6}
	\end{figure}

    Fig. \ref{Fig6} gives more examples of the spectra in which the noise is removed based on self-correlation of the data, thus avoiding the limitation and negative effect of the training set. It can be seen that all the de-noised spectra have high SNR levels and the energy band features are more visible, indicating that the self-correlation information of a single spectrum is sufficient for extracting the noise. These further confirm the universality and versatility of our training-set-free method.

\section*{\label{sec:level2} APPENDIX B: Implementation Details}
	Parameters for the neural network architecture are listed in Table \ref{Tab:arch}, where BN stands for batch normalization \cite{Ioffe2015} and the parameter of Leaky ReLU (lReLU) is set to $0.2$. We use the bilinear upsampling to increase the feature size in the decoder. It has been shown \cite{Ulyanov2020,You2020} that the performance is not very sensitive to the choice of the input, so we follow these practices in this work.

\begin{table}[H]
\caption{Neural network architecture}\label{Tab:arch}
\setlength\tabcolsep{7pt} 
\begin{tabular}{ccc}
  \hline\noalign{\smallskip}
  Encoder network\\\hline
  Input spectra $x$ \\
  Conv2d, BN, $3 \times 3 \times 16$, stride=2, lReLU \\
  Conv2d, BN, $3 \times 3 \times 32$, stride=2, lReLU \\
  Conv2d, BN, $3 \times 3 \times 64$, stride=2, lReLU \\
  Conv2d, BN, $3 \times 3 \times 128$, stride=2, lReLU \\
  Conv2d, BN, $3 \times 3 \times 128$, stride=2, lReLU \\
  \hline Decoder network\\\hline
  Conv2d, BN, $3 \times 3 \times 128$, stride=1, lReLU \\
  Upsampling, Conv2d, BN, $3 \times 3 \times 128$, stride=1, lReLU \\
  Upsampling, Conv2d, BN, $3 \times 3 \times 64$, stride=1, lReLU \\   
  Upsampling, Conv2d, BN, $3 \times 3 \times 32$, stride=1, lReLU \\
  Upsampling, Conv2d, BN, $3 \times 3 \times 16$, stride=1, lReLU \\
  Conv2d, $1 \times 1 \times 1$, stride=1, sigmoid\\
  \hline\noalign{\smallskip}
\end{tabular}
\end{table}

\section*{\label{sec:level3} APPENDIX C: The Role of Depth}
\begin{figure}[H]
    \centering
    \includegraphics[width=0.48\textwidth]{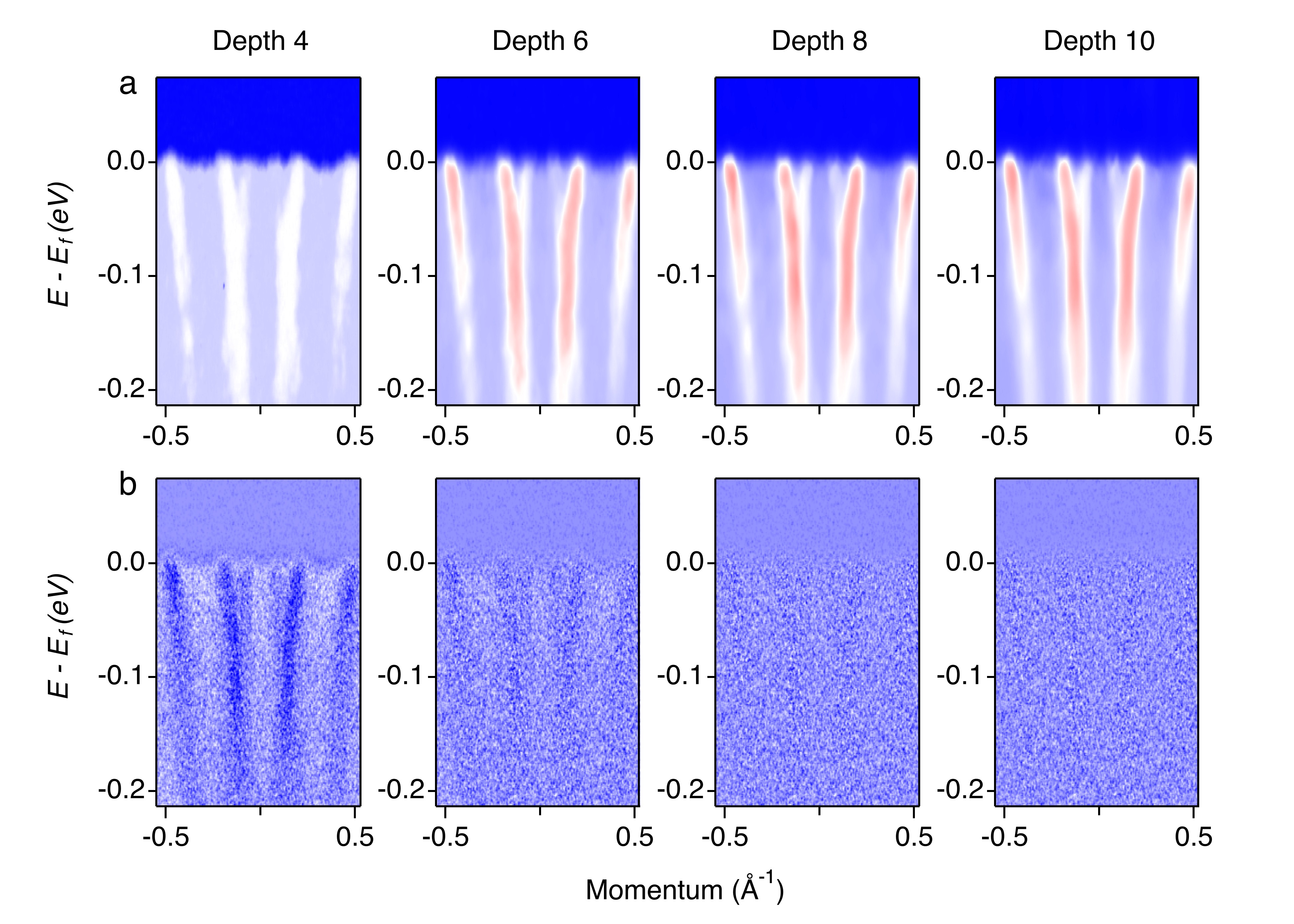}
    \caption{Depth dependent de-noised results and predicted noise structures. (a) De-noised results of the cuprate ARPES spectra with varying depth 4, 6, 8, 10. (b) Corresponding noise structures of (a).}
    \label{Fig7}
\end{figure}

Depth of neutral network plays a crucial role in deep learning. To illustrate the effect of depth in our algorithm, we have also tested 8, 6, and 4 layer U-Net with the same other hyperparameters as the 10 layer U-Net. The depth dependent de-noised results are shown in Fig. \ref{Fig7}. We see that depth indeed influences the performance of our algorithm. Neural network with a depth larger than 4 can already capture main features of the intrinsic energy band. Slightly reducing the depth may worsen the results, but it is still possible to use a shallower neural network. However, if the neural network is not deep enough, it cannot produce clean spectra with correct energy band information. There exists a threshold determining the success of our algorithm.  When the depth is lower than the threshold, the algorithm fails and cannot extract the complete intrinsic information, leaving part of the energy band signal in the noise structure, as may be seen from the results of depth 4 in Fig. \ref{Fig7}. With the increase of depth, the output spectra become more and more smooth and satisfactory. Therefore, we should use a neural network with depth at least larger than this threshold in practice.

\hspace{0.1cm}


\begin{thebibliography}{100}

\bibitem{Damascelli2003} A. Damascelli, Z. Hussain, and Z.-X. Shen, Rev. Mod. Phys. {\bf 75}, 473 (2003).

\bibitem{Lv2021} B. Q. Lv, T. Qian, and H. Ding, Rev. Mod. Phys. {\bf 93}, 025002 (2021).

\bibitem{Sobota2021} J. A. Sobota, Y. He, and Z.-X. Shen, Rev. Mod. Phys. {\bf 93}, 025006 (2021).

\bibitem{Binnig1983} G. Binnig, and H. Rohrer, Surf. Sci. {\bf 126}, 236 (1983).

\bibitem{Pan2001a} S. H. Pan, J. P. O'Neal, R. L. Badzey, C. Chamon, H. Ding, J. R. Engelbrecht, Z. Wang, H. Eisaki, S. Uchida, A. K. Gupta, K.-W. Ng, E. W. Hudson, K. M. Lang, and J. C. Davis, Nature. {\bf 413}, 282 (2001).

\bibitem{Chaix2013a} L. Chaix, S. de Brion, F. L\'evy-Bertrand, V. Simonet, R. Ballou, B. Canals, P. Lejay, J. B. Brubach, G. Creff, F. Willaert, P. Roy, and A. Cano, Phys. Rev. Lett. {\bf 110}, 157208 (2013).
 
\bibitem{Kotani2001a} A. Kotani, and S. Shin, Rev. Mod. Phys. {\bf 73}, 203 (2001).

\bibitem{Medjanik2017a} K. Medjanik, O. Fedchenko, S. Chernov, D. Kutnyakhov, M. Ellguth, A. Oelsner, B. Sch\"onhense, T. R. F. Peixoto, P. Lutz, C.-H. Min, F. Reinert, S. D\"aster, Y. Acremann, J. Viefhaus, W. Wurth, H. J. Elmers, and G. Sch\"onhense, Nature Mater. {\bf 16}, 615 (2017).

\bibitem{Zhou2005a} X. J. Zhou, B. Wannberg, W. L. Yang, V. Brouet, Z. Sun, J. F. Douglas, D. Dessau, Z. Hussain, and Z.-X. Shen, J. Electron Spectrosc. Relat. Phenom. {\bf 142}, 27 (2005).

\bibitem{Graf2010a} J. Graf, S. Hellmann, C. Jozwiak, C. L. Smallwood, Z. Hussain, R. A. Kaindl, L. Kipp, K. Rossnagel, and A. Lanzara, J. Appl. Phys. {\bf 107}, 014912 (2010).

\bibitem{Smallwood2012b} C. L. Smallwood, C. Jozwiak, W. T. Zhang, and A. Lanzara, Rev. Sci. Instrum. {\bf 83}, 123904 (2012).

\bibitem{He2016a} Y. He, I. M. Vishik, M. Yi, S. Yang, Z. Liu, J. J. Lee, S. Chen, S. N. Rebec, D. Leuenberger, A. Zong, C. M. Jefferson, R. G. Moore, P. S. Kirchmann, A. J. Merriam, and Z.-X. Shen, Rev. Sci. Instrum. {\bf 87}, 011301 (2016).

\bibitem{Mills2019a} A. K. Mills, S. Zhdanovich, M. X. Na, F. Boschini, E. Razzoli, M. Michiardi, A. Sheyerman, M. Schneider, T. J. Hammond, V. S\"{u}ss, C. Felser, A. Damascelli, and D. J. Jones, Rev. Sci. Instrum. {\bf 90}, 083001 (2019).

\bibitem{Grass2010a} M. E. Grass, P. G. Karlsson, F. Aksoy, M. Lundqvist, B. Wannberg, B. S. Mun, Z. Hussain, and Z. Liu, Rev. Sci. Instrum. {\bf 81}, 053106 (2010).

\bibitem{Nitta2019a} J. Nitta, K. Miwa, N. Komiya, E. Annese, J. Fujii, S. Ono, and K. Sakamoto, Sci. Rep. {\bf 9}, 9645 (2019).

\bibitem{Peng2020a} H. Peng, X. Gao, Y. He, Y. Li, Y. Ji, C. Liu, S. A. Ekahana, D. Pei, Z. Liu, Z. Shen, and Y. Chen, Rev. Sci. Instrum. {\bf 91}, 033905 (2020).

\bibitem{Kim2021a} Y. Kim, D. Oh, S. Huh, D. Song, S. Jeong, J. Kwon, M. Kim, D. Kim, H. Ryu, J. Jung, W. Kyung, B. Sohn, S. Lee, J. Hyun, Y. Lee, Y. Kim, and C. Kim, Rev. Sci. Instrum. {\bf 92}, 073901 (2021).

\bibitem{Huang2022} D. Huang and Y.-F. Yang, Phys. Rev. B. {\bf 105}, 075112 (2022).

\bibitem{Dong2021} J.-J. Dong, D. Huang, and Y.-F. Yang, Phys. Rev. B. {\bf 104}, L081115 (2021).

\bibitem{Antun2020} V. Antun, F. Renna, C. Poon, B. Adcock, and A. C. Hansen, Proc. Natl. Acad. Sci. U.S.A. {\bf 117}, 30088 (2020).

\bibitem{Bhadra2021} S. Bhadra, V. A. Kelkar, F. J. Brooks, and M. A. Anastasio, IEEE T Med Imaging. {\bf 40}, 3249 (2021).

\bibitem{Li2013a} S. Li, H. Yin, and L. Fang, IEEE Trans. Geosci. Remote Sens. {\bf 51}, 4779 (2013).

\bibitem{Ulyanov2020} D. Ulyanov, A. Vedaldi, and V. Lempitsky, Int. J. Comput. Vis. {\bf 128}, 1867 (2020).

\bibitem{Krizhevsky2017} A. Krizhevsky, I. Sutskever, and G. E. Hinton, Commun. Acm. {\bf 60}, 84 (2017).
 
\bibitem{Brown2020} T. Brown, B. Mann, N. Ryder, M. Subbiah, J. D. Kaplan, P. Dhariwal, A. Neelakantan, P. Shyam, G. Sastry, A. Askell, S. Agarwal, A. Herbert-Voss, G. Krueger, T. Henighan, R. Child, A. Ramesh, D. Ziegler, J. Wu, C. Winter, C. Hesse, M. Chen, E. Sigler, M. Litwin, S. Gray, B. Chess, J. Clark, C. Berner, S. McCandlish, A. Radford, I. Sutskever, and D. Amodei, in Advances in Neural Information Processing Systems, edited by H. Larochelle, M. Ranzato, R. Hadsell, M.F. Balcan, and H. Lin (Curran Associates, Inc., San Francisco, 2020), pp. 1877-1901. 

\bibitem{Ronneberger2015UNet} O. Ronneberger, P. Fischer, and T. Brox, in Medical Image Computing and
Computer-Assisted Intervention -- MICCAI 2015, edited by Nassir Navab, Joachim Hornegger, William~M. Wells, and
Alejandro~F. Frangi (Springer International Publishing, Cham, 2015), pp. 234-241.

\bibitem{You2020} C. You, Z. Zhu, Q. Qu, and Y. Ma, in Advances in Neural Information Processing Systems, edited by H. Larochelle, M. Ranzato, R. Hadsell, M.F. Balcan, and H. Lin (Curran Associates, Inc., San Francisco, 2020), pp. 17733-17744.

\bibitem{Paszke2019} A. Paszke, S. Gross, F. Massa, A. Lerer, J. Bradbury, G. Chanan, T. Killeen, Z. Lin, N. Gimelshein, L. Antiga, A. Desmaison, A. Kopf, E. Yang, Z. DeVito, M. Raison, A. Tejani, S. Chilamkurthy, B. Steiner, L. Fang, J. Bai, and S. Chintala, in Advances in Neural Information Processing Systems, edited by H. Wallach, H. Larochelle, A. Beygelzimer, F. d'Alch\'{e}-Buc, E. Fox and R. Garnett (Curran Associates, Inc., San Francisco, 2019).

\bibitem{Candes2011} E.~J. Cand\`{e}s, X.~Li, Y.~Ma, and J.~Wright, {J. ACM} \textbf{58}, 1 (2011).

\bibitem{Ioffe2015} S. Ioffe and C. Szegedy, in Proceedings of the 32nd International Conference on Machine Learning, edited by Francis Bach and David Blei (PMLR, Lille, 2015), pp. 448-456.

\end{thebibliography}
\end{document}